\documentclass[twocolumn, amssymb, amsmath, nobibnotes, nofootinbib, aps, prb,showpacs]{revtex4-1}
\usepackage[dvips]{graphicx}
\begin{document}
\title{Short range magnetic correlation, metamagnetism and coincident dielectric anomaly in Na$_5$Co$_{15.5}$Te$_6$O$_{36}$}
\author{Rafikul Ali Saha$^1$}
\author{Jhuma Sannigrahi$^2$}
\author{Ilaria Carlomagno$^3$}
\author{Somdatta Kaushik$^4$}
\author{Carlo Meneghini$^3$}
\author{Mitsuru Itoh$^5$}
\author{Vasudeva Siruguri$^4$}
\author{Sugata Ray$^{1}$}
\email{mssr@iacs.res.in}
\affiliation{$^1$School of Materials Sciences, Indian Association for the Cultivation of Science, 2A \& 2B Raja S. C. Mullick Road, Jadavpur, Kolkata 700 032, India}
\affiliation{$^2$School of Physical Sciences, Indian Institute of Technology Goa, Goa-403401, India}
\affiliation{$^3$Dipartimento di Scienze, Universitá Roma Tre, Via della Vasca Navale, 84 I-00146 Roma, Italy}
\affiliation{$^4$UGC-DAE-Consortium for Scientific Research Mumbai Centre, 246C 2nd floor Common Facility Building (CFB), Bhabha Atomic Research Centre, Mumbai 400085, India}
\affiliation{$^5$Materials and Structures Laboratory, Tokyo Institute of Technology, 4259 Nagatsuta, Yokohama 226-8503, Japan}

\pacs {}
\begin{abstract}
Here we explore the structural, magnetic and dielectric properties of Co based compound Na$_5$Co$_{15.5}$Te$_6$O$_{36}$ as a candidate of short-range magnetic correlations driven development of dielectric anomaly above N$\acute{e}$el temperature of ($T_N$=) 50 K. Low temperature neutron powder diffraction (NPD) in zero applied magnetic field clearly indicates that the canted spin structure is responsible for the antiferromagnetic transition and partially occupied Co form short range magnetic correlation with other Co, which further facilitates the structural distortion and consequent development of dielectric anomaly above antiferromagnetic transition. Additionally, the temperature dependent magnetic heat capacity and electron spin resonance measurements reveal the presence of short-range magnetic correlations which coincides with an anomaly in the dielectric constant vs temperature curve. Moreover, significant changes in the lattice parameters are also observed around the same temperature, indicating presence of noticeable spin-lattice coupling. Further, sharp jump in the magnetic field dependent magnetization clearly indicates the presence of metamagnetic transition and magnetic field dependent NPD confirms that rotations of Co spins with applied magnetic field are responsible for this metamagnetic phase transition. As a result, this transition causes the magnetocaloric effect to be developed in the system, which is suitable for the application in low temperature refrigeration.
\end{abstract}
\maketitle
\section{Introduction}
The exploration of multiferroic materials have been the focus of extensive research in recent years due to their potential applications in electronic devices like Spintronic devices, Information storage devices, Spin valve, Quantum electro magnets, Microelectronic devices, Sensors etc. as well as their importance on fundamental science understanding~\cite{Tokura, Spaldin}. Based on the microscopic origin of ferroelectricity, multiferroic materials are classified into two different groups, namely, proper and improper multiferroics. In case of proper multiferroics, the sources of ferroelectricity and magnetism are different and they respond independently, though some coupling between them may be present. The spontaneous electric polarization in these materials ($P$) is often large ($\sim$10-100 $\mu$C/cm$^2$)~\cite{Dasgupta, Khomskii}. There are several mechanisms for developing ferroelectricity in this class of multiferroioc materials, such as i) ferroelectricity due to lone pair (PbTiO$_3$, Pb$_3$TeMn$_3$P$_2$O$_{14}$, BiMnO$_3$, BiFeO$_3$, SnTiO$_3$ etc.)~\cite{Jin, Rafikul-PRB, Hill, Volkova, Nakhmanson, Seshadri}, ii) ferroelectricity due to charge ordering (Pr$_{0.5}$Ca$_{0.5}$MnO$_3$, RMnO$_3$, TbMn$_2$O$_5$, LuFe$_2$O$_4$)~\cite{Khomskii2, Mostovoy, Hur, Ikeda}, iii) geometric ferroelectricity (YMnO$_3$)~\cite{Spaldin2} etc. But in case of improper multiferroicity, ferroelectricty appears due to the development of particular spin order which breaks the lattice inversion symmetry through exchange striction, thereby inducing a small amount of spontaneous polarization ($\sim$10$^{-2}$ $\mu$C/cm$^2$)~\cite{Dasgupta, Khomskii}. In these cases, external magnetic field influences the arrangement of magnetic structure, generally leading to changes in dielectric properties. Due to the strong magnetoelectric coupling (ME), improper multiferroic materials are more important from the application point of view. Both noncollinear (TbMnO$_3$, Ni$_3$V$_2$O$_6$, MnWO$_4$)~\cite{Khomskii, Balatsky, Mostovoy2} and collinear (Ca$_3$CoMnO$_6$)~\cite{Choi} magnetic structures can generate ferroelectricity in the system. Spiral magnetic order, resulting from magnetic frustration, develop exchange striction which is associated with the antisymmetric part of the exchange coupling and constitutes Dzyaloshinsky-Moriya (DM) interaction~\cite{Balatsky, Mostovoy2, Dagotto, Wohlman}. In case of collinear magnetic order, lattice relaxation through exchange striction is associated with the symmetric superexchange coupling~\cite{Choi, Cheong}.
\par
Surprisingly, short-range magnetic correlation driven multifferoicity has been less explored till date. There are few compounds such as Ca$_3$Co$_2$O$_6$, Er$_2$BaNiO$_5$ which show displacive-type ferroelectricity involving off-centering of the magnetic ion due to short-range magnetic correlation~\cite{Sampathkumaran1, Sampathkumaran2}. Another langasite compound Pb$_3$TeMn$_3$P$_2$O$_{14}$ which shows covalency driven modulation of paramagnetism and development of ferroelectricity~\cite{Rafikul-PRB}. Co based compounds are important in this context, where possible existence of different Co-oxidation states (+2, +3 and +4), with several local coordinations (tetrahedra, octahedra, trigonal prism and so on) around Co and most importantly various spin states within same Co valance often offer intriguing physical properties~\cite{Sampathkumaran2, Paul, Kalobaran}. Therefore we choose one Co based compound Na$_5$Co$_{15.5}$Te$_6$O$_{36}$~\cite{Shan}, where two types of local coordinations (octahedra and trigonal prism) of single valance Co (2+) are present and the super-exchange interaction pathways within edge shared Co$^{2+}$ octahedras are around 90$^\circ$, while corner connected Co-O-Co bond angles are around 120$^\circ$ which signify the presence of two types of magnetic interactions. Another interesting thing is that here some Co and Na are distributed  inhomogeneously because they are occupying same position with occupancy 0.63 and 0.37 for Na and Co respectively. This gives rise to local inhomogeneity and as a result, we may expect that this compound would possess only short-range magnetic correlations.
\par
Here we have discussed the crystal structure, magnetic structure and dielectric properties of the system. Low temperature magnetic structure (from Neutron powder diffraction) revels the presence of canted antiferromagnetic ordering. In addition to the antiferromagnetic transition ($T_N$), the $dc$ magnetic susceptibility measurements further suggest the presence of short range ferromagnetic correlation above $T_N$, while an anomaly is observed at the same temperature in temperature dependent dielectric data as well as lattice parameter variations, indicating the presence of magnetoelectric coupling in the system. Further electron spin resonance and magnetic heat capacity support the presence of short-range magnetic correlation above the N$\acute{e}$el temperature (50 K). The system also exhibits a metamagnetic transition in magnetic field dependent isothermal magnetization measurement, inducing the magnetocaloric effect in the system, which is suitable for the application in low temperature refrigeration. In addition, magnetic field dependent NPD confirms that rotations of Co spins with applied magnetic field are responsible for this metamagnetic phase transition.
\section{Methodology}
\subsection{Experimental techniques}
Polycristalline sample Na$_5$Co$_{15.5}$Te$_6$O$_{36}$ (NCTO) was prepared using conventional solid state reaction technique. NCTO sample were synthesized by taking stoichiometric amounts of high purity Na$_2$CO$_3$ (Merck $\geq$ 99 \%), TeO$_2$ (Sigma-Aldrich 99.9995 \%) and Co$_3$O$_4$ (Sigma-Aldrich 99.99 \%). The mixture was initially calcined at 600$^{\circ}$~C in air for 12 hour. The resultant mixture was then reground and finally annealed at 800$^{\circ}$~C for  12 hour in air. The phase purity of the sample was checked from X-ray powder diffraction (XRD) measured at Bruker AXS: D8 Advanced x-ray diffarctometer equipped with Cu $K$$_{\alpha}$ radiation. Temperature dependent XRD were carried out at RIGAKU Smartlab (9KW) XG equipped with Cu $K_{\alpha}$ to realize the presence of temperatutre dependent structural phase transitions. The obtained XRD data were analyzed using Rietveld technique and refinements of the crystal structure were performed by FULLPROF program ~\cite{Carvajal}. Neutron powder diffraction (NPD) measurements were carried out on powder samples using the multiposition sensitive detector based focussing crystal diffractometer set up by UGC-DAE Consortium for Scientific Research Mumbai Centre at the National Facility for Neutron Beam Research (NFNBR), Dhruva reactor, Mumbai (India). NPD at room temperature and 3 K have been taken using a wavelength of 2.315 \AA, while field dependent NPD at 10 K have been performed using a wavelength of 1.48 \AA. The NPD data were analyzed via refinement using the JANA2006 program~\cite{Palatinus}. Co $K$-edge x-ray absorption fine-structure (XAFS) spectra were collected at the XAFS beamline of the Elettra Synchrotron Centre, Italy. The $dc$ Magnetic measurements were carried out using a superconducting quantum interference device (SQUID) magnetometer (Quantum Design, USA) over a temperature range: 2-300 K in magnetic fields upto $\pm$ 7 Tesla and also in a vibrating-sample magnetometer (Cryogenic, UK). For calculation of magnetocaloric effect, virgin curve of magnetic field dependent magnetization ($M-H$) upto 16 T have been taken using vibrating sample magnetometer (VSM) probe in a cryogenic  physical property measurement system (PPMS). The heat capacity was measured by relaxation method in a quantum design  physical property measurement system (PPMS). The permitivities were measured using a Hewlett-PacPrecision LCR meter (HP4284A) at an $ac$ level of 1 V mm$^{-1}$. A cryogenic temperature system (Niki Glass LTS-250-TL-4W) was used to control the temperature within the range of 4 - 350 K. The dielectric hysteresis loops were measured using a ferroelectric measurement system (Toyo Corporation FCE-3) equipped with an Iwatsu ST-3541 capacitive displacement meter having a linearity of 0.1\% and a resolution of 0.3 nm. X-band Electron paramagnetic resonance (EPR) measurements were performed on a Jeol JES-200 spectrometer.
\section {Results and Discussions:}
\subsection{Structural and Electronic Characterization}
Powder x-ray diffraction (XRD) data of Na$_5$Co$_{15.5}$Te$_6$O$_{36}$ (NCTO) at 300 K confirms phase purity and the structural refinement has been performed by considering a hexagonal space group $P$6$_3$/m, which is consistent with previous literature report~\cite{Shan}, as shown in Fig. 1(a) and Table I. Temperature dependent X-ray diffractions of NCTO compound have been performed over a wide temperature range of 5-300 K. The Rietveld refinements of these collected XRD patterns have been done using the same space group $P$6$_3$/m. Further, room temperature neutron powder diffraction (NPD) of the sample has been carried out to get exact position and occupancy of lighter oxygen atoms, as shown in Fig. 1(b). Similar result like XRD was obtained from the refinement of the powder neutron diffraction data at room temperature. The crystal structure of  NCTO contains one formula per unit-cell, depicted in Fig. 1(c). Here, both Te and Co are coordinated by six oxygen ions but geometry of their polyhedra are different. Te and Co(1) form a slightly distorted octahedra, while two Co (Co(2) and Co(3)) make distorted trigonal prisms. Both room temperature XRD and NPD reveals the presence of rare disorder between Co(3) and Na(3), where site occupancy of Co(3) and Na(3) are 0.37 and 0.63 respectively. Among the Co polyhedra, Co(3)/Na(3) polyhedra is larger due to the greater ionic radius of Na$^+$. Two nearest neighbor Co(3)/Na(3) distorted trigonal prisms are connected via common face sharing with each others, while Co(3)/Na(3) prism is attached with Co(1) octahedra and Co(2) trigonal prism via edge sharing and common face sharing respectively. On the other hand, two Co(1) octahedra are connected via edge sharing and one Co(1) octahedra is connected with Co(2) trigonal prism via corner sharing.
\begin{figure}
%\centering
\resizebox{8.6cm}{!}
{\includegraphics[86pt,287pt][488pt,710pt]{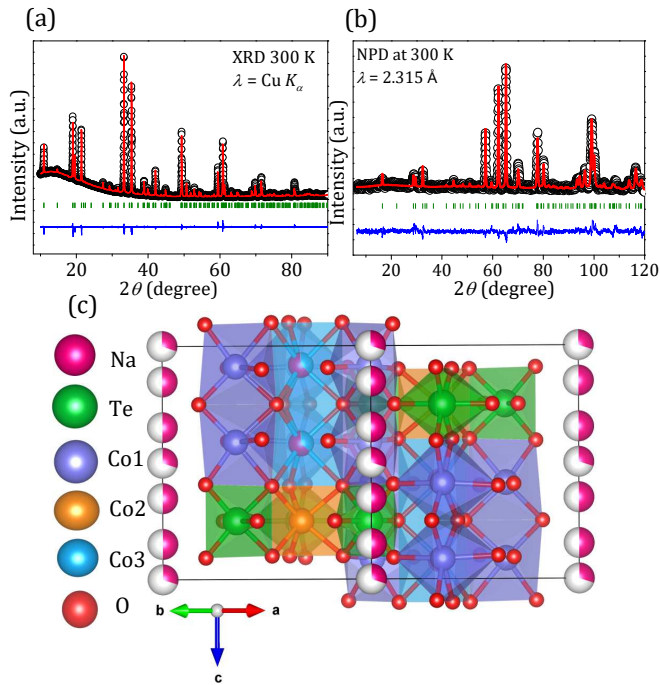}}
\caption{(a)and (b) Rietveld refined XRD and NPD of NCTO at room temperature respectively. Black open circles represent the experimental data and red line represents the calculated pattern. The blue line represents the difference between the observed and calculated pattern and green lines signify the position of Bragg peaks. (c) Refined crystal structure of NCTO.}
%\end{figure*}
\end{figure}

\begin{table}
%\begin{center}
\caption{The sample NCTO is refined within a single crystallographic phase using a hexagonal space group $P$6$_3$/m. Lattice parameters of NCTO: $a$ = $b$ = 9.3231(9) {\AA}, $c$ = 9.0682(7) {\AA}, $\alpha$ = $\beta$ = 90$^{\circ}$, $\gamma$ = 120$^{\circ}$. $R_{wp}$ = 14.51, $R_{exp}$ = 8.16, $\chi^2$ = 3.16.}
\resizebox{9cm}{!}{
\begin{tabular}{ c  c  c  c  c  c }\hline
Atom & $x$(\AA)  &  $y$(\AA) & $z$(\AA) & $B_{iso}$ & Occupancy\\\hline
Te &  0.6560(9) &  0.0098(5) & 0.0000 & 0.142(8) & 1.0\\
Co(1) &  0.3438(1) &  0.9934(5) &  0.0877(9) & 0.213(5) & 1.0\\
Co(2) &  0.66667 &  0.33333 &  0.2500 & 0.213(5) & 1.0\\
Na(1) & 0.0000	& 0.0000 & 0.0000 & 0.258(4) & 0.3\\
Na(2) & 0.0000	& 0.0000 & 0.6520(6) & 0.258(4) & 0.5\\
Co(3) & 0.33333 & 0.66667 & 0.0818(3) & 0.213(5) & 0.37\\
Na(3) & 0.33333 & 0.66667 & 0.0818(3) & 0.213(5) & 0.63\\
O(1) &  0.6904(4) &  0.8883(8) &  0.0908(6) & 0.174(3) & 1.0\\
O(2) &  0.6097(7) &  0.1400(2) &  0.1007(4) & 0.174(3) & 1.0\\
O(3) &  0.8805(1) &  0.1732(6) &  0.2500 & 0.174(3) & 1.0\\
O(4) &  0.4132(8) &  0.8627(3) &  0.2500 & 0.174(3) & 1.0\\
\hline
\end{tabular}
}
%\end{center}
\end{table}
\par
In order to maintain the charge neutralization of the sample, the oxidation state of Na, Co and Te should be 1+, 2+ and 6+ respectively. To know the valance state of the cations we have carried out bond valance sum calculation (BVS) using the formula $V$ = $\sum$ exp(($R_0$-$R_i$)/b), where $V$ is the valance state of an cation, $R_i$ is the observed bond length, $R_0$ is the tabulated parameter expressing the ideal bond length when the element $i$ has exactly valence 1 and b is an empirical constant, typically 0.37 \AA, shown in Table II. Te, Co(1) and Co(2) show 5.8, 1.9 and 1.8 oxidation state respectively but surprisingly partially occupied Co(3) (with Na(3)) shows 1.2 valency which is much smaller than usual 2+, while 1.1 valency for Na. Therefore in order to accommodate Co(3) at the Na(3) site, the Co-O bond length should be changed and consequently a distortion would be developed in the Co(3) containing trigonal prism polyhedra.
\begin{table}
\caption{Bond valance sum calculation.}
\resizebox{6cm}{!}{
\begin{tabular}{| c | c | c |}
\hline Element & Bond length (\AA) &  valency \\\hline
 Co(1) & 2*Co(1)-O(1) - 2.05 & 1.9 \\
    & 2*Co(1)-O(2) - 2.24 &   \\
    & Co(1)-O(3) - 2.07 &   \\
     & Co(1)-O(4) - 2.18 &   \\
 Co(2) & 6*Co(2)-O(2) - 2.14 & 1.8 \\
 Co(3) & 3*Co(3)-O(2) - 2.34 & 1.2 \\
    & 3*Co(3)-O(4) - 2.24 &   \\
 Na(3) & 3*Na(3)-O(2) - 2.34 & 1.1 \\
    & 3*Na(3)-O(4) - 2.24 &   \\
 Te & 2*Te-O(1) - 1.91 & 5.8 \\
     & 2*Te-O(2) - 1.97 &   \\
    & Te-O(3) - 1.90 &   \\
     & Te-O(4) - 1.95 &   \\
 \hline
\end{tabular}
}
\end{table}
\par
As Co always has the tendency to be in a variety of charge states as well as spin states in the Co based oxide compounds~\cite{Korotin, Komarek, Tjeng}, precise estimation of Co oxidation state in the present compound is of central importance in the context of Co mediated physical properties. Therefore, to know the true valency of Co, Co $K$-edge X-ray absorbtion near edge spectroscopy (XANES) has been carried out and presented in Fig. 2(a). Consequently, the first order derivative of the normalized XANES spectrum (see inset of Fig. 2(a)) identifies a pre edge peak at 7708.9 eV (Due to 1$s$ to 3$d$ transition caused by 3$d$-4$p$ mixing or dipole allowed 1$s$ to 4$p$ transition~\cite{Hambley, Glatzel}) and an absorption edge peak at 7717.6 eV (Due to transition Co 1$s$ orbital to molecular orbital with Co 4$p$~\cite{Hambley, Glatzel}). We have compared the XANES of the experimental data (NCTO sample) with those from CoO (being Co$^{2+}$) and CoO(OH) (being full Co$^{3+}$), as shown in Fig. 2(a). The raising of the edge on the data perfectly matches with the Co$^{2+}$ state (see Fig. 2(a)), which signifies the 2+ oxidation state of Co in the sample~\cite{CoO1, CoO2, CoOOH}.
\begin{figure}
%\centering
\resizebox{8.6cm}{!}
{\includegraphics[55pt,398pt][418pt,737pt]{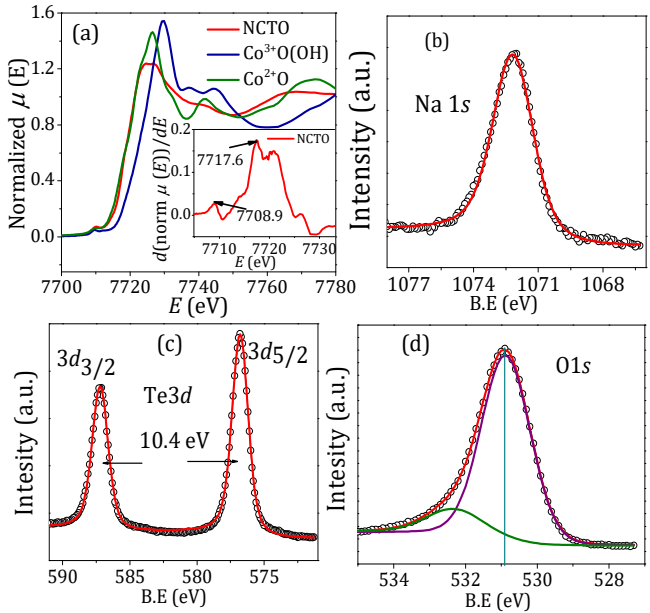}}
\caption{(a) Normalized Co $K$-edge X-ray absorption near edge spectra of NCTO (red curve), CoO(OH)(blue curve) and CoO (green curve). Inset shows the first order derivative of normalized data. XPS spectra of (b) Na 1$s$, (c)Te 3$d$,  and (d) O 1$s$. Black open circles and red lines are the experimental and fitted data respectively. Green and violet line are the two singlets for O 1$s$.}
%\end{figure*}
\end{figure}
\par
Further x-ray photoelectron spectroscopy (XPS) measurements were carried out to confirm the oxidation states of the constituent elements of this compound. The core level Na 1$s$, Te 3$d$ and O 1$s$ spectra, along with their reasonable fitting (shown in Figs. 2(b)-(d)) have clearly indicated the respective charge states. The energy position of the Na 1$s$ (1072.1 eV) clearly reveals the expected 1+ charge state~\cite{Na-XPS}, while the peak positions of Te 3$d$ doublets (576.8 eV, 587.2 eV) along with their energy separation (10.4 eV) suggest 6+ valance state of Te in NCTO~\cite{Te}. The O 1$s$  spectrum of NCTO has been fitted by considering two singlets (see the green and violet line of Fig. 2(d)), which signifies the presence of the multiple oxygen site in this compound~\cite{Bandyopadhyay}.
\subsection{Magnetic, Thermodynamic and Dielectric Characterization}
Thermal variations of magnetic susceptibility ($\chi$) have been measured under $H$ = 100 and 500 Oe in the zero-field-cooled (ZFC), field-cool-cooling (FCC), and field-cool-heating (FCH) conditions. ZFC and FCH at 100 Oe have been shown in Fig. 3(a). A clear antiferromagnetic transitions ($T_N$) around 50 K is observed in both the ZFC and FC curve, consistent with previous study~\cite{Shan}. Curie-Weiss (C-W) fit ($\chi$ = $\chi_0$ + $C$/($T$ - $\theta_{\text{CW}}$); where $\chi_0$ is the temperature independent paramagnetic succeptibility, $C$ the Curie constant related to the effective moment, and $\theta_{\text{CW}}$ the Weiss temperature) on the 100 Oe field cooled heating susceptibility data in the temperature range $T$= 180 - 350 K gives the negative Weiss temperature ($\theta_{\text{CW}}$ = -6.3 K), which indicates the antiferromagnetic interactions between the Co spins. Effective paramagnetic moment $\mu_{\text{eff}}$ = 3.9 $\mu_{\text{B}}$/Co, obtained from the C-W analysis, resemble quite well with the theoretical value 3.87 $\mu_{\text{B}}$/Co for high spin Co$^{2+}$. Interestingly, the C-W fit deviates from the experimental 1/$\chi$ ($T$) versus $T$ curve, in the form a sharp downturn, from around much higher temperature of 155 K (see Fig. 3(b)), which is further supported by the respective first order derivative curves (see inset of Fig. 3(b)). This kind of signature is usually attributed to the Griffiths phase singularity~\cite{Griffiths, Bray}, where short-range (SR) FM clusters~\cite{Tokura2} appear upon cooling, or equivalently, finite spin-canted clusters get sustained by a strong basal-plane anisotropy in microscopically phase separated regions much above the bulk $T_N$~\cite{Basheed}. A prominent dip in first temperature derivative of magnetic susceptibility data (see inset of Fig. 3(a)) also supports the presence of ferromagnetic interaction above N$\acute{e}$el temperature of this compound.
\begin{figure}
%\centering
\resizebox{8.6cm}{!}
{\includegraphics[67pt,94pt][537pt,747pt]{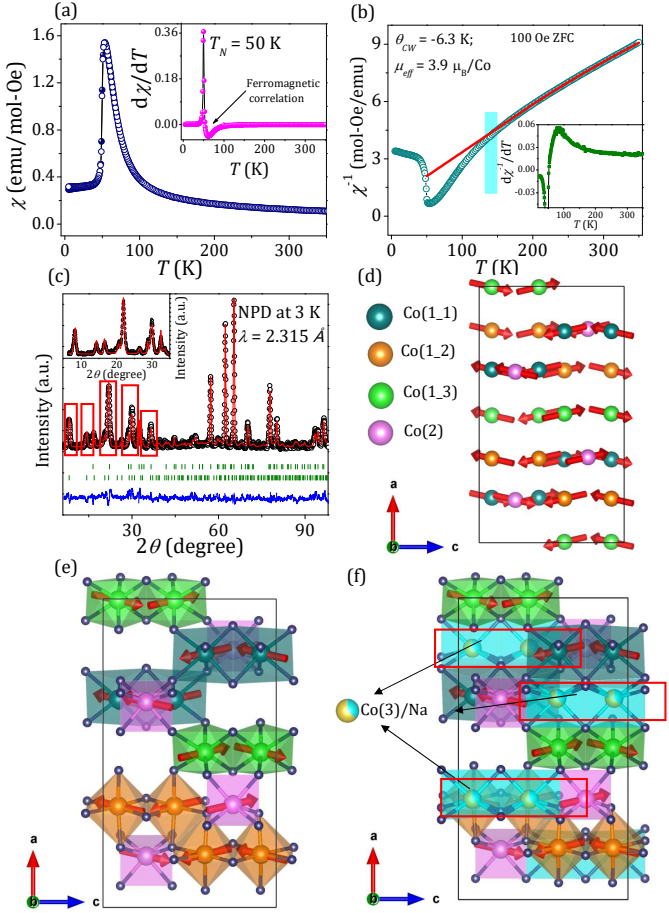}}
\caption{(a) ZFC and FCH of NCTO at $H$ = 100 Oe. Inset shows the first order derivative of $\chi (T)$. (b) Inverse susceptibility data at 100 Oe. Open circle data and red line are experimental and Curie-Weiss fitted data respectively. Inset shows the first order derivative of $\chi^{-1} (T)$. (c) Rietveld refined NPD of NCTO at 3 K. Black open circles represent the experimental data and red line represents the calculated pattern. The blue line represents the difference between the observed and calculated pattern and green lines signify the position of Bragg peaks. (d) Refined magnetic structure of NCTO. (e) All Co polyhedra connections with spin structure which are responsible for long range ordering. (f) Partially occupied Co polyhedra and spin structure of fully occupied Co. Red squares are the positions of partially occupied Co (Co3/Na).}
%\end{figure*}
\end{figure}
\par
In order to further clarify the magnetic ground state of NCTO, we have conducted neutron powder diffraction (NPD) at room temperature (see Fig. 1(b)) and below long range ordering temperature (3 K) (shown in Fig. 3(c)). Low temperature pattern reveals few additional resolutions limited Bragg peaks (forbidden in the structural space group) as well as an increase of the intensity of some nuclear peaks (see the red rectangle box and inset of Fig. 3(c)). The additional intensities are observed only at a lower 2-theta region pointing to their magnetic origin. All the magnetic reflections can be indexed using the $k$ =(0.5 0.5 0) propagation vector. The magnetic structure has been refined using JANA2006 software considering the Shubnikov magnetic space group $P$[$B$]21/m, origin shifted by (1/4, 0, 1/4). We have performed the magnetic structure refinement without considering long range ordering at the partially occupied Co3 site (Co/Na = 0.37/0.63) because the Co(3) moments share same site with Na, and random presence of Na at the Co(3) site hinders the Co(3) ions from participating in the long range magnetic interactions with the other two Co-sites. Within this assumption we get a reliable fitting which shows a canted antiferromagnetic structure where the spins of fully occupied Co(1-1), Co(1-2), Co(1-3), and Co(2) are aligned along $c$ axis but canted towards $a-b$ plane (see Fig. 3(d)). Magnetic moment per Co site ($\mu_{eff}$ = 3.7 $\mu_B$/Co), obtained from neutron refinement, is almost resembles well to the spin only moment of Co$^{2+}$ for high-spin configuration (3.87 $\mu_B$/Co). Refinement of 3 K data shows that Co spins within the edge shared Co motifs (two orange and green octahedra) are directed along $c$ axis but canted within $a-b$ plane, as depicted in Fig. 3(e). These connectivities satisfy the Co-O-Co bond angle $\sim$ 84.8$^\circ$ $\sim$ 96.1$^\circ$ of the super-exchange pathways. On the other hand, spins belonging to the 1-$D$ zig-zag chains of the corner shared CoO$_6$ octahedra (see orange and green octahedra in Fig. 3(e)) with the Co-O-Co bond angle of 123.5$^\circ$, cause mutually opposite canted spin alignment. The CoO$_6$ triangular prisms (pink polyhedra in Fig. 3(e)) also share their corners with the neighbouring Co motifs creating the bond angles of 106.1$^\circ$ and 155.7$^\circ$. These canted antiferromagnetic spin structures induce antiferromagnetic long range ordering at 50 K. However, another interesting incident appears when partially occupied Co(3) is introduced in the spin structure, as shown in Fig. 3(f). Partial occupancy between Co(3) and Na signifies that Co(3) is distributed inhomogeneously in the unit cell. As a result, where Co(3) is present in the unit cell, it is connected via face sharing with Co(2) trigonal prism, edge sharing with Co(1-1), Co(1-2) and Co(1-3), as shown in Fig. 3(f). The downward deviation from C-W law in the inverse susceptibility data, indicating the presence of short range magnetic correlation, may therefore be attributed to the magnetic interactions of partially occupied Co(3) with another neighboring Co ions.

\begin{figure}
%\centering
\resizebox{8.6cm}{!}
{\includegraphics[64pt,548pt][429pt,715pt]{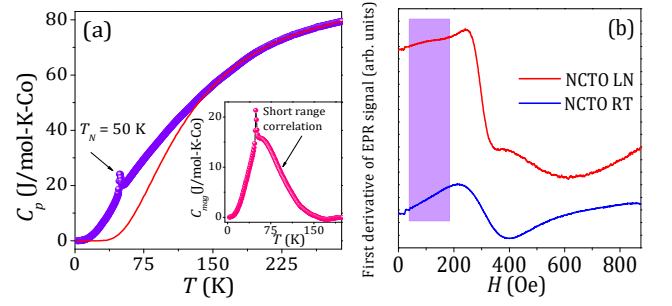}}
\caption{(a) Heat capacity data (violet) and Debye-Einstein fitting (red line) and inset shows magnetic part of heat capacity of NCTO. (b) EPR spectra at room temperature (blue line) and liquid nitrogen temperature (red line).}
%\end{figure*}
\end{figure}
\par
The magnetic transitions were further established by the zero field heat capacity measurements ($C_p$ versus $T$), as shown in Fig. 4(a). A sharp lambda-like anomaly has been observed near 50 K, indicating the antiferromagnetic transition and it is in agreement with magnetic susceptibility data. The high temperature part of the $C_p$($T$) data has been fitted with a combination of Debye and Einstein functions of heat capacity, $C_p$($T$) = \`{D}($\Theta_D$, $T$)+$\sum_{i=1}^{2}$ $\xi$($\Theta_{Ei}$, $T$), where \`{D} and $\xi$ are Debye and Einstein functions, respectively~\cite{Tomy}. Lattice part in the heat capacity data ($C_{latt}$) was determined by extrapolating the fitted data to low temperature the fitted data in absence of suitable nonmagnetic sample. The magnetic contribution to the heat capacity ($C_{mag}$) is obtained by subtracting the lattice component from the total heat capacity data. Interestingly, a clear hump like behaviour appears above 50 K in the $C_{mag}$ versus $T$ curve, shown in inset of Fig.  4(a), which indicates the presence of short range magnetic correlations above $T_N$.

\par
To further identify the presence of ferromagnetic clusters in paramagnetic phase of the NCTO sample, we have performed the electron paramagnetic resonance (EPR) measurements. The variations of first-derivative EPR spectra ($dP$/$dH$) in the paramagnetic temperature region (300 K and liquid nitrogen temperature) are shown in Fig. 4(b). A single symmetric line without any structural change was observed in room temperature EPR data (blue line) which signified a typical behaviour for a normal paramagnetic phase at room temperature. Compared to this room temperature paramagnetic resonance line, the extra broad hump (highlighted portion) has been observed in the liquid nitrogen temperature data (red line) which indicates the presence of some localized ferrromagnetic interactions~\cite{EPR1, EPR2, EPR3}.
\begin{figure}
%\centering
\resizebox{8.6cm}{!}
{\includegraphics[41pt,419pt][501pt,771pt]{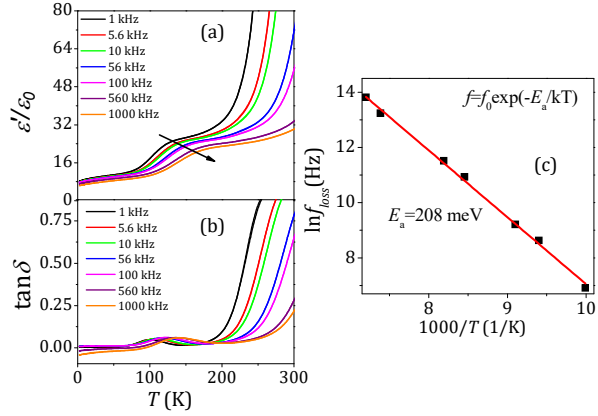}}
\caption{(a) and (b) Temperature dependence of real part of dielectric constant $\varepsilon'$/$\varepsilon_0$ and $\tan\delta$ loss data of NCTO at different frequencies. (c) Shows the fitting of the activation energy corresponding to the dielectric loss.}
%\end{figure*}
\end{figure}

\begin{figure}
%\centering
\resizebox{8.6cm}{!}
{\includegraphics[123pt,260pt][457pt,785pt]{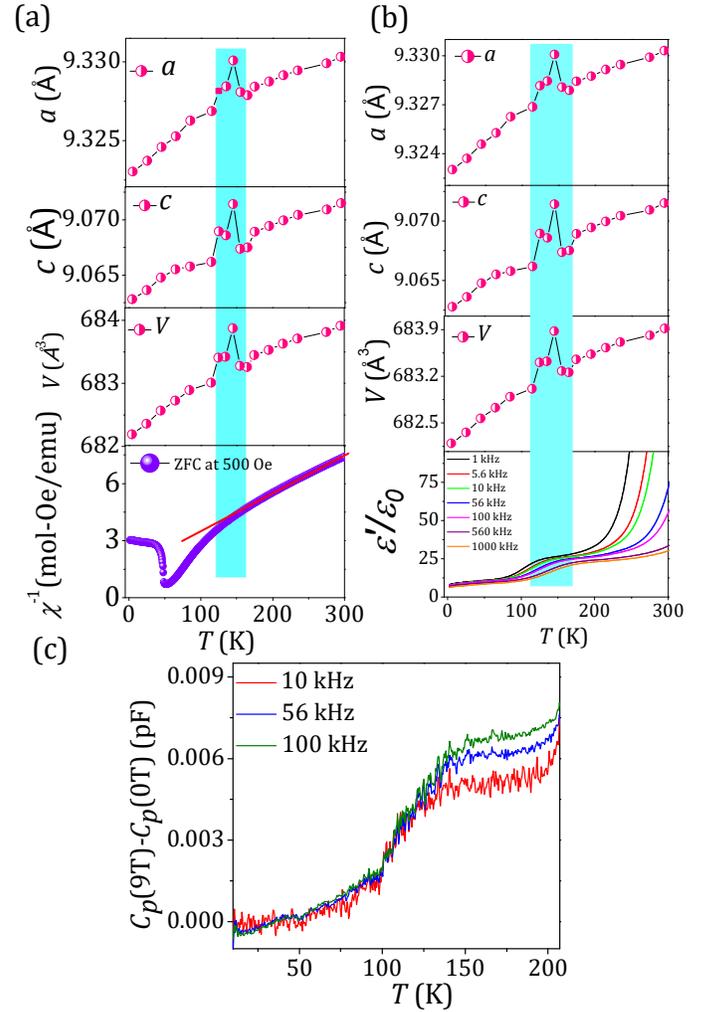}}
\caption{(a) Temperature dependence of lattice parameters, unit cell volume and inverse susceptibility data. (b) Thermal variation of real part of dielectric constant $\varepsilon'$/$\varepsilon_0$ and lattice parameters. (c) Difference between the thermal variation of capacitance at applied magnetic field 9 T and zero field.}
%\end{figure*}
\end{figure}
\par
Both cooling and heating cycle of the temperature dependence of the dielectric constant ($\varepsilon'$/$\varepsilon_0$) and dielectric loss ($\tan\delta$) of NCTO at frequencies from 1 kHz - 1000 kHz in the temperature range of 2 to 300 K are shown in Figs. 5(a) and (b) respectively. A glass like phase transition appears near 155 K in the both temperature dependent dielectric constant and loss ($\tan\delta$) data, which shift to higher temperature with increasing frequencies implying a thermally activated relaxation process. $f_{loss}$ is the frequency corresponding to the peak of the loss (tan$\delta$) curves at the corresponding temperature $T$ which are plotted as $\ln f$ versus 1000/$T$ plot, as shown in Fig. 5(c). The relaxation shows linear behaviour which can be fitted to the Arrhenius law $f$ = $f_0$$\exp(E_a/k_BT)$, where $E_a$ is the activation energy, $f_0$ is the pre-exponential factor and $k_B$ is the Boltzmann constant, the fitting reveals the thermal activation energy ($E_a$) of relaxation to be 208 meV. A sharp anomaly near 155 K in thermal variations of refined lattice parameters $a$, $b$, $c$ and volume of the unit cell ($V$) coincide with temperature dependence of the dielectric anomaly, as shown in Fig. 6(b), signifying the correlation between structural distortion and dielectric anomaly. Due to rather tiny changes in position coordinates across the phase transition, the present Rietveld refinements could not conclusively determine the true symmetry of low temperature phase. Such difficulties are not unusual when the lattice distortions are weak~\cite{Rafikul-PRB, Sampathkumaran1}. However, it is interesting to note that the observed anomalies in $a$, $b$, $c$, and volume at around 155 K coincide with the magnetic and dielectric anomalies, as can be observed in Fig. 6(a). Further we study the temperature dependent capacitance under 9 Tesla magnetic field and zero field and plot the difference (see Fig. 6(c)). A very small but significant change is observed between the with field ($H$ = 9T) and without field data near $\sim$155 K, as indicated in Fig. 6(c), signifying the presence of magnetoelectric coupling in the system. As a result, we can conclude that structural distortion is responsible for dielectric anomaly. So, we may comment that short range ferromagnetic correlations develop among the Co(3) spins at around 155 K which is simultaneously related Co(3) off-centering and dielectric anomaly at the same temperature.
\begin{figure}
%\centering
\resizebox{8.6cm}{!}
{\includegraphics[45pt,193pt][531pt,790pt]{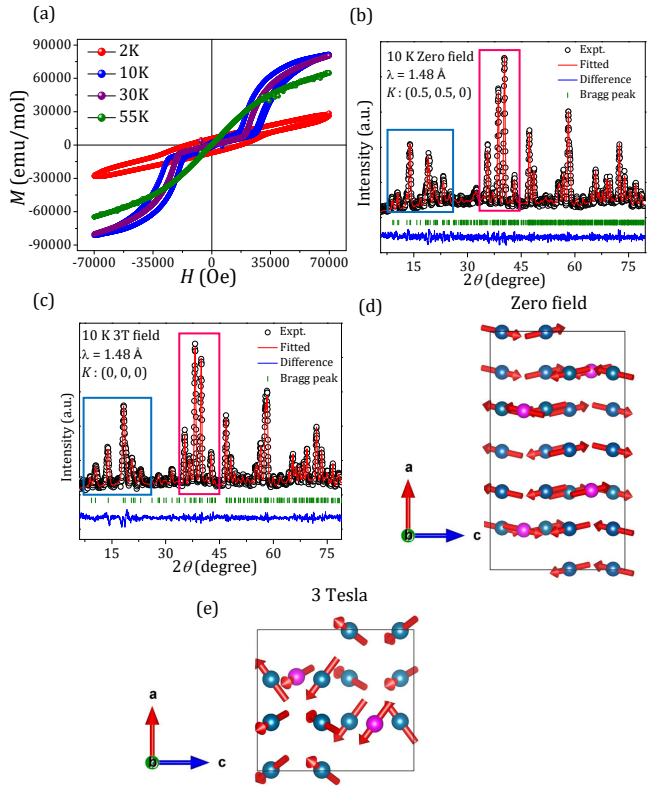}}
\caption{(a) $M$ versus $H$ data at 2, 10, 30 and 55 K. (b) and (c) Refined neutron powder diffraction pattern in magnetic field $\mu_0H$ = 0 T and 3 T at 10 K, respectively. Black open circles represent the experimental data and red line represents the calculated pattern. The blue line represents the difference between the observed and calculated pattern and green lines signify the position of Bragg and magnetic peaks. (d) and (e) magnetic structure of NCTO compound in different magnetic field $\mu_0H$ = 0 T and 3 T, respectively. Dark cyan and pink spheres are Co(1) and Co(2), respectively.}
%\end{figure*}
\end{figure}
\subsection{Field induce change in magnetic response}
Further, we have measured isothermal magnetization data ($M$-$H$) at different temperature (2, 10, 30 and 55 K), as shown in Fig. 7(a). The $M$-$H$ curve at 2 K exhibits a high coercive field ($H_C$ = 25 kOe) and non-saturating behaviour up to 7 Tesla, indicating the presence of high magnetocrystalline anisotropy within the predominantly antiferromagnetic interaction. Surprisingly, a sudden jump around 26.5 kOe of applied field in the 10 K $M$-$H$ data signifies the presence of metamagnetic phase transition. The signature of metamagnetism is also present at 30 K $M-H$ data, but at slightly lower magnetic field (see Fig. 7(a)). Additionally, this compound shows almost $\sim$ 9\% magneto-caloric effect (MCE) at applied 16 T magnetic field (Detailed calculation has been given in Supplementary material).
\par
Furthermore, we have carried out the neutron powder diffraction under magnetic fields $\mu_0H$ = 0 T and 3 T at 10 K to ascertain the modification in spin structure beyond the metamagnetic phase transition. Neutron powder diffraction data at 3 K and 10 K have been performed using neutron beam of 2.315 \AA~and 1.48 \AA, respectively. The reason for choosing higher wavelength neutron beam was to look for any additional magnetic peak in the low angle region.
\par
NPD at 10 K for zero and 3 T applied magnetic field collected at 1.48 \AA~ have been depicted in Figs. 7(b)-(c). Clear differences are observed in the NPD diffraction pattern with increasing magnetic field pointing towards certain change in the  magnetic unit cell as well as spin structure with applied magnetic field. In order to unfold the origin of all the observed field induced subtle changes in the NPD data we have first fitted the zero field data of 10 K using the propagation vector [0.5, 0.5, 0] and the Shubnikov magnetic space group $P$[$B$]21/m, origin shifted by (1/4, 0, 1/4), which gave the similar kind of spin structure to the zero filed 3 K data. Interestingly, attempts to fit the 3 T data using the same propagation vector [0.5, 0.5, 0] as zero field failed to capture all the extra peaks generated under the externally applied magnetic fields. On the other hand, systematic changes in the recorded magnetic peak intensities with increasing applied magnetic field have been quite well captured by assuming the [0, 0, 0] propagation vector. The magnetic structures of the 3 T data have been refined considering the Shubnikov magnetic space group $P$63$'$/$m$, as shown in Figs. 7(c), and got quite satisfactory fitting. Consequently, the magnetic unit cell and the respective spin structures (obtained from the refinement) for zero and 3 T applied magnetic field have been illustrated in Figs. 7(d)-(e). The depicted field-induced changes in the spin structures (see Figs. 7(d)-(e)) clearly demonstrate the spin-flop transition mechanism upon applied magnetic fields to be responsible for the observed metamagnetic transition in $M-H$ curve (see Fig. 7(a)) of this system. The refined values of the Co magnetic moments are 3.8 and 3.9 $\mu_B$ for the zero and 3 T field respectively, which clearly indicates that moment value increases with increasing applied magnetic field.

\section{Conclusion}
We have reported the results of structural, magnetic and dielectric properties in details. XRD and NPD reveal that among the Co polyhedra, partially occupied Co(3) becomes larger due to the greater ionic radius of Na$^+$. According to BVS calculation, this polyhedra does not satisfy the valency of that Co(2+). Therefore, to maintain the proper oxidation state of Co, local symmetry of the Co(3) polyhedra should be modified. From the low temperature magnetic spin structure, we can say that partially occupied Co create a ferromagnetic cluster in the system which help to change the polyhedra of that Co. Magnetic heatcapacity and EPR data also say the presence of short range magnetic correlation above N\'{e}el temperature. Therefore, an anomaly due to short range magnetic correlation coincides with the temperature dependent lattice parameter  as well as dielectric anomaly. A very small but significant changes in between the with field ($H$ = 9T) and without field data near $\sim$155 K signify the presence of magnetoelectric coupling in the system. As a result, we may conclusively state that the appearance of magnetoelectric coupling in this compound aries from the local off centering of Co(3) centres, caused due to the developed short range magnetic correlations among the Co(3) spins. In addition, isothermal magnetization as well as magnetic field dependent NPD clearly affirm that spin flop transition with increasing magnetic field causes metamagnetic phase transition. However, our results introduce another short range magnetic correlation driven multiferroic system as well as a spin flop metamagnetic candidate.

\section{Acknowledgement}
RAS thanks CSIR, India for a fellowship. SR thanks Technical Research Center of IACS. SR also thanks Department of Science and Technology (DST) [Project No. WTI/2K15/74], UGC-DAE Consortium for Research, Mumbai, India [Project No. CRS-M-286] for support.


\begin{thebibliography}{99}
\bibitem{Tokura} T. Kimura, T. Goto, H. Shintani, K. Ishizaka, T. Arima, and Y. Tokura, Nature (London) {\bf426}, 55 (2003).
\bibitem{Spaldin} N. A. Spaldin and M. Fiebig, Science {\bf309}, 391 (2005).
\bibitem{Dasgupta}  J. Chakraborty, N. Ganguli, T. Saha-Dasgupta and I. Dasgupta, Phys. Rev. B {\bf88}, 094409 (2013).
\bibitem{Khomskii} D. Khomskii Physics {\bf2}, 20 (2009).
\bibitem{Jin} X. He and K. Jin, Phys. Rev. B {\bf94}, 224107 (2016).
\bibitem{Rafikul-PRB} R. A. Saha, A. Halder, T. Saha-Dasgupta, D. Fu, M. Itoh, and S. Ray, Phys. Rev. B {\bf101}, 180406(R) (2020).
\bibitem{Hill} N. A. Hill and K. M. Rabe, Phys. Rev. B {\bf59}, 8759 (1999).
\bibitem{Volkova} L. M. Volkova and D. V. Marinin, J. Supercond. Nov. Magn. {\bf24}, 2161-2177 (2011).
\bibitem{Nakhmanson} K. C. Pitike, W. D. Parker, L. Louis and S. M. Nakhmanson, Phys. Rev. B {\bf91}, 035112 (2013).
\bibitem{Seshadri} R. Seshadri and N. A. Hill, Chem. Mater {\bf13}, 2892-2899 (2001).
\bibitem{Khomskii2} D. V. Efremov, J. van den Brink, and D. I. Khomskii, Nature Mater. {\bf3}, 853 (2004).
\bibitem{Mostovoy} S. W. Cheong and M. V. Mostovoy, Nature Mater. {\bf6}, 13 (2007).
\bibitem{Hur} N. Hur, S. Park, P. A. Sharma, J. S. Ahn, S. Guha and S-W. Cheong, Nature {\bf429}, 392 (2004).
\bibitem{Ikeda} N. Ikeda, H. Ohsumi, K. Ohwada, K. Ishii, T. Inami, K. Kakurai,Y. Murakami, K. Yoshii, S. Mori, Y. Horibe and H. Kito, Nature
{\bf436}, 1136 (2005).
\bibitem{Spaldin2} B. B. van Aken, T. T. M. Palstra, A. Filippetti and N. A. Spaldin, Nature Mater. {\bf3}, 164 (2004).
\bibitem{Balatsky} H. Katsura, N. Nagaosa and A. V. Balatsky, Phys. Rev. Lett. {\bf95}, 057205 (2005).
\bibitem{Mostovoy2} M. V. Mostovoy, Phys. Rev. Lett. {\bf96}, 067601 (2006).
\bibitem{Choi} Y. J. Choi, H. T. Yi, S. Lee, Q. Huang, V. Kiryukhin and S.-W. Cheong, Phys. Rev. Lett. {\bf100}, 047601 (2008).
\bibitem{Dagotto} I. A. Sergienko and E. Dagotto, Phys. Rev. B {\bf73}, 094434 (2006).
\bibitem{Wohlman} A. B. Harris, T. Yildirim, A. Aharony and O. Entin-Wohlman, Phys. Rev. B {\bf73}, 184433 (2006).
\bibitem{Cheong} L. C. Chapon, P. G. Radaelli, G. R. Blake, S. Park, and S.-W. Cheong, Phys. Rev. Lett. {\bf96}, 097601 (2006).
\bibitem{Sampathkumaran1} T. Basu, V. V. R. Kishore, S. Gohil, K. Singh, N. Mohapatra, S. Bhattacharjee, B. Gonde, N. P. Lalla, P. Mahadevan, S. Ghosh and E. V. Sampathkumaran, Sci. Rep. {\bf4}, 5636 (2014).
\bibitem{Sampathkumaran2} T. Basu, K. K. Iyer, K. Singh and E. V. Sampathkumaran, Sci. Rep. {\bf3}, 3104 (2013).
\bibitem{Paul} V. Hardy, S. Lambert,M. R. Lees and D. M. Paul, Phys. Rev. B {\bf68}, 014424 (2003).
\bibitem{Kalobaran} R. Bindu, K. Maiti, S. Khalid and E. V. Sampathkumaran, Phys. Rev. B {\bf79}, 094103 (2009).
\bibitem{Shan} Y. J. Shan, Y. Yoshioka, M. Wakeshima, K. Tezuka and H. Imoto, J. Solid State Chem. {\bf211}, 63$–$68 (2014).
%\bibitem{Pico} l. Alvarez, M. L. Veiga and C. Pico, J. Mater. Chem. {\bf5(7)}, 1049$-$1051 (1995).
%\bibitem{Carbonio} M.S. Augsburger, M.C.Viola, J.C.Pedregosa, A.Munoz, J.A.Alonso and R.E. Carbonio, J. Mater. Chem. {\bf15}, 993$–$1001 (2005).
%\bibitem{Itoh} D. Iwanaga, Y. Inaguma, M. Itoh, J. Solid State Chem. {\bf147}, 7291$–$295 (1999).
%\bibitem{Kleinke} J. Xu, A. Assoud, N. Soheilnia, S. Derakhshan, H. L. Cuthbert, J. E. Greedan, M. H. Whangbo, H. Kleinke, Inorg. Chem. {\bf44(14)}, 5042–5046 (2005).
%\bibitem{Cava} L. Viciu, Q. Huang, E. Morosan, H. W. Zandbergen, N. I. Greenbaum, T. McQueen, R. J. Cava, J. Solid State Chem. {\bf180}, 1060–1067 (2007).
%\bibitem{Cussen} M. P. O’Callaghan, A. S. Powell, J. J. Titman, G. Z. Chen, E. J. Cussen, Chem. Mater. {\bf20}, 2360$–$2369 (2008).
\bibitem{Carvajal}  J. Rodriguez Carvajal, Physica B {\bf192}, 55 (1993).
\bibitem{Palatinus} V. Petricek, M. Dusek, and L. Palatinus, Z. Kristallogr. - Cryst. Mater. B {\bf229} 345 (2014).
\bibitem{Korotin}  M. A. Korotin, S. Yu. Ezhov, I. V. Solovyev, V. I. Anisimov, D. I. Khomskii and G. A. Sawatzky Phys. Rev. B {\bf54} 5309 (1996).
\bibitem{Komarek} Z. W. Li, Y. Drees, C. Y. Kuo, H. Guo, A. Ricci, D. Lamago, O. Sobolev, U. Rütt, O. Gutowski, T. W. Pi, A. Piovano, W. Schmidt, K. Mogare, Z. Hu, L. H. Tjeng and A. C. Komarek, Sci. Rep. {\bf6}, 25117 (2016).
\bibitem{Tjeng}J. Chen, Y. Chin, M. Valldor, Z. Hu, J. Lee, S. Haw, N. Hiraoka, H. Ishii, C. Pao, K. Tsuei, J. Lee, H. Lin, L. Jang, A. Tanaka, C. Chen, and L. H. Tjeng, J. Am. Chem. Soc.{\bf136}, 1514$-$1519 (2014).
\bibitem{Hambley} P. D. Bonnitcha, M. D. Hall, C. K. Underwood, G. J. Foran, M. Zhang, P. J. Beale and T. W. Hambley, J. Inorg. Biochem.{\bf100}, 963$-$971 (2006)
\bibitem{Glatzel} F. de Groot, G. Vank\'{o} and P. Glatzel, J. Phys.: Condens. Mater. {\bf21}, 104207 (2009).
\bibitem{CoO1} B. You, N. Jiang, M. Sheng, S. Gul, J. Yano, and Y. Sun, Chem. Mater. {\bf27}, 7636-7642 (2015).
\bibitem{CoO2} H. C. Choi, S. Y. Lee, S. B. Kim, M. G. Kim, M. K. Lee, H. J. Shin, and J. S. Lee, J. Phys. Chem. B {\bf106}, 9252-9260 (2002).
\bibitem{CoOOH} J. Zhou, Y. Wang, X. Su, S. Gu, R. Liu, Y. Huang, S. Yan, J. Li and S. Zhang, Energy Environ. Sci., {\bf12}, 739 (2019).
\bibitem{Na-XPS} K. J. Gaskell, A. L. Asunskis, and P. M. A. Sherwood, Surface Science Spectra {\bf9}, 151 (2002).
\bibitem{Te} P. Pal, A. Sahoo, Md. F. Abdullah, S. D. Kaushik, P. N. Vishwakarma, and A. K. Singh, J. Appl. Phys.{\bf124}, 164110 (2018).
\bibitem{Bandyopadhyay} A. Bandyopadhyay, S. K. Neogi, A. Paul, C. Meneghini, I. Dasgupta, S. Bandyopadhyay and S. Ray, Phys. Rev. B {\bf95} 024432 (2017).
\bibitem{Griffiths}  R.B. Griffiths, Phys. Rev. Lett. {\bf23} 17 (1969).
\bibitem{Bray}  A. J. Bray, Phys. Rev. Lett. {\bf59} 586$-$589 (1987).
\bibitem{Tokura2}  J. Deisenhofer, D. Braak, H.-A. Krug von Nidda, J. Hemberger, R. M. Eremina, V. A. Ivanshin, A. M. Balbashov, G. Jug, A. Loidl, T. Kimura, Y. Tokura, Phys. Rev. Lett. {\bf95} 257202 (2005).
\bibitem{Basheed}  A. Rathia, P.K. Rout, Sonam Perweena, R.P. Singh, P.D. Babu, Anurag Guptaa, R.P. Panta, G.A. Basheed, J. Magn. Magn. Mater.{\bf468} 230-234 (2018).
\bibitem{Tomy}  R. P. Singh and C. V. Tomy, J. Phys.: Condens. Mater. {\bf20}, 235209 (2008).
\bibitem{EPR1}  J. Deisenhofer, D. Braak, H.-A. Krug von Nidda, J. Hemberger, R. M. Eremina, V. A. Ivanshin, A. M. Balbashov, G. Jug, A. Loidl, T. Kimura and Y. Tokura, Phys. Rev. Lett. {\bf95}, 257202 (2005).
\bibitem{EPR2}  S Angappane, M Pattabiraman, G Rangarajan, K Sethupathi, Babu Varghese and V S Sastry, J. Phys.: Condens. Matter {\bf19}, 036207 (2007).
\bibitem{EPR3}  L. Chen, J. Fan, W. Tong, D. Hu, Y. Ji, J. Liu, L. Zhang, L. Pi, Y. Zhang and H. Yang, Sci. Rep. {\bf6}, 14 (2016).
%\bibitem{Siruguri} A. K. Singh, S. Patnaik, S. D. Kaushik, and V. Siruguri, Phys. Rev. B {\bf81}, 184406 (2010).
\end{thebibliography}
\end{document}